\documentclass[journal=jacsat,manuscript=article]{achemso}

\usepackage[version=3]{mhchem} 
\usepackage{graphicx}
\usepackage{dcolumn}
\usepackage{bm}
\usepackage[T1]{fontenc}
\usepackage{esvect}
\usepackage{amsbsy}
\usepackage{amssymb}
\usepackage{amsmath}
\usepackage[makeroom]{cancel}
\usepackage{graphicx}
\usepackage{grffile}
\graphicspath{ {./images/} }
\usepackage{caption}
\usepackage{subcaption}
\usepackage{float}
\usepackage{diffcoeff}
\usepackage{physics}
\usepackage[colorinlistoftodos]{todonotes}
\usepackage[colorlinks=true, allcolors=blue]{hyperref}
\usepackage{chemformula}
\usepackage{indentfirst}
\usepackage{parskip}
\usepackage{color}
\usepackage{comment}
\usepackage{mathrsfs}
\usepackage{slashed}
\usepackage{hyphenat}
\usepackage{natbib}
\author{Shuo Tao}
\affiliation{Department of Physics, Rutgers University, Newark, NJ 07102, United States of America}
\author{Xuecheng Shao}
\affiliation{Department of Physics, Rutgers University, Newark, NJ 07102, United States of America}
\author{Li Zhu}
\affiliation{Department of Physics, Rutgers University, Newark, NJ 07102, United States of America}
\email{li.zhu@rutgers.edu}

\title{Accelerating Structural Optimization through Fingerprinting Space Integration on the Potential Energy Surface}

\abbreviations{DFT, PES, PSO, GA, AI, GTOs, MLPs, EPs, GOM, SCM, PCA}

\begin{document}







\begin{abstract}

Structural optimization has been a crucial component in computational materials research, and structure predictions have relied heavily on this technique in particular. In this study, we introduce a novel method that enhances the efficiency of local optimization by integrating an extra fingerprint space into the optimization process. Our approach utilizes a mixed energy concept in the hyper potential energy surface (PES), combining real energy and a newly introduced fingerprint energy derived from the symmetry of local atomic environment. 
This method strategically guides the optimization process toward high-symmetry, low-energy structures by leveraging the intrinsic symmetry of atomic configurations.
The effectiveness of our approach was demonstrated through structural optimizations of silicon, silicon carbide, and Lennard-Jones cluster systems. Our results show that the fingerprint space biasing technique significantly enhances the performance and probability of discovering energetically favorable, high-symmetry structures, as compared to conventional optimizations. The proposed method is anticipated to streamline the search for new materials and facilitate the discovery of novel, energetically favorable configurations.

\end{abstract}



The design and discovery of novel materials has been a cornerstone in the modern technological advancements, where computational materials science emerges as a critical pillar, driving forward the innovative quest for new materials. 
Central to this computational endeavor is the prediction of the structure of materials, a fundamental information in understanding their properties and functionalities at the atomic scale.
Structure prediction methods, such as CALYPSO\cite{wangCrystalStructurePrediction2010,WANG20122063}, USPEX\cite{GLASS2006713}, XtalOpt\cite{LONIE2011372}, basin hopping\cite{walesGlobalOptimizationBasinHopping1997a}, minima hopping\cite{goedeckerMinimaHoppingEfficient2004a}, and random structure search\cite{pickardInitioRandomStructure2011}, therefore, play a critical role in navigating this challenge, forming the backbone of material exploration and discovery. 
To predict a new (meta-)stable material, we need to study its thermodynamic stability with respect to numerous local minima on a high-dimensional potential energy space (PES) in the first place. However, finding the most thermodynamically stable structure of a large assembly of atoms is a very difficult problem, because the number of minima on the PES of a large system increases exponentially with the number of atoms \cite{PhysRevE.59.48,DC9766100012,doi:10.1021/j100145a019}.

To surmount this obstacle, several methods have been developed to improve the capabilities of structure searching methods. One approach is to introduce a bias into the search space, targeting the high symmetry structures that are more likely to be the global minimum. It is extensively studied in both crystal \cite{doi:10.1021/ja01379a006} and cluster \cite{walesSymmetryNearsymmetryEnergetics1998} systems that structures with high symmetry tend to have either very low or very high energy (Pauling’s ``rule of parsimony''). This may explain why most of the ground state structures that exist in nature have high symmetry. It would be feasible if one could take advantage of this symmetry bias during the local optimization process and hence reduce the computation time to investigate the local minima residing in the high energy funnels. For example, Shao \textit{et al.} \cite{shaoSymmetryorientatedDivideandconquerMethod2022} introduced a symmetry tree graph coupled with an artificial-intelligence-based symmetry selection strategy, markedly simplifying the problem of crystal structure prediction by bypassing the exploration of the intricate low-symmetry subspace within the entire search space. A recent work \cite{huberTargetingHighSymmetry2023} shows that by explicitly introducing a symmetry biased penalty function to the PES, high-symmetry structures will be found much faster than on an unbiased surface when performing structure predictions using the minima hopping method.

While these global optimization methods vary in their specific structure searching approaches, they all rely on the integrative calling of the local optimization process. However, local optimization methods often involve lengthy computation times and high levels of complexity. Numerous local energy minima, many of which have high energy levels, can impede the optimization process. This impediment is especially noticeable in strong covalent systems like carbon and silicon. Once covalent bonds are formed, they pose significant barriers to reconfiguration. Thus, even if more energetically favorable structures are within reach, accessing them remains challenging, trapping the system in a potentially energy-intensive state. This situation has sparked interest in more efficient methods capable of accelerating this process. For example, by starting a structure stochastically generated in a higher-dimensional space (hyperspace), the structure can be relaxed in the additional dimension, which  has been shown to be effective in enhance the probability of reaching low-energy configurations \cite{pickardHyperspatialOptimizationStructures2019}.

In this work, we introduce a new method to enhance the efficiency of local optimization by introducing an extra symmetry space. Rather than starting the structural optimizations initiated from a structure stochastically generated in hyperspace, we propose a new implicit approach to infuse symmetry information into the process of local optimization. This strategy, designed to prevent configurations from stagnating in high-energy states, is realized through the introduction of the fingerprint energy that servers as the indicator of the symmetry of structures. 
This performance boost is anticipated to be advantageous for PES exploration methods that rely on the local optimization of structures. Therefore, the work provides a path toward the objective of predicting the material structure of complex systems.

In our approach, we introduce the fingerprint space to define the mixed energy ($E_{\text{mixed}}$) on the hyper PES as follows:
\begin{align}
&E_{\text{mixed}}=\omega_{\text{fp}}E_{\text{fp}}+\omega_{\text{real}}E_{\text{real}},
\end{align}
where $\omega_{\text{fp}}$ and $\omega_{\text{real}}$ are the mixing weights, and $E_{\text{real}}$ represents the real energy on the normal PES, as described by methods such as density functional theory (DFT), machine-learning potentials, or empirical potentials. $E_{\text{fp}}$ is the fingerprint energy on the fingerprint PES, which is defined based on the sum of the Euclidean norm of all pairwise atomic fingerprint vector\cite{zhuFingerprintBasedMetric2016} distances within the crystal structure:
\begin{align}
&E_{\text{fp}} = \eta\sum_{i=1}^{N_{\text{at}}}\sum_{j>i}^{N_{\text{at}}}\left\|\mathbf{fp}_{i}-\mathbf{fp}_{j}\right\|^{2},
\end{align}
where $N_{\text{at}}$ denotes the number of atoms in the unit cell, and $\mathbf{fp}_{i}$ and $\mathbf{fp}_{j}$ are the atomic fingerprint vectors for the atoms indexed by $i$ and $j$, respectively. The coefficient $\eta$ is a scaling factor utilized to equate the units of our fingerprint energy to the corresponding physical units. The atomic fingerprint vectors ($\mathbf{fp}_{i}$) are employed to characterize the chemical environment of atom ($i$) within structures. These vectors are derived from the eigenvalues of the localized Gaussian overlap matrix\cite{10.1063/1.4828704,zhuFingerprintBasedMetric2016}, a method proven to be both efficient and reliable in distinguishing various atomic environments.\cite{zhuFingerprintBasedMetric2016,parsaeifardFingerprintBasedDetectionNonLocal2021,parsaeifardManifoldsQuasiconstantSOAP2022}

The concept of symmetry in crystal structures is intimately linked to the similarity of local environments of atoms within the structure. In high-symmetry structures, atoms typically share similar local environments, reflecting a uniformity and orderliness in their spatial arrangement. Conversely, low-symmetry structures are characterized by a diversity in the local environments of atoms, indicating a more irregular and varied atomic arrangement. This contrast in local environments serves as a fundamental indicator of the overall symmetry of a structure.
The degree of similarity in the local environments of different atoms can be effectively quantified to gauge the symmetry of a structure. Here, atomic fingerprints emerges as an efficient tool. By assessing the differences in atomic fingerprints across a structure to obtain a direct measure of its symmetry, the fingerprint energy becomes an effective metric.
Lower fingerprint energy corresponds to higher structural symmetry, indicating a more uniform distribution of atomic environments throughout the crystal. A structure where all atomic environments are identical, such as in a diamond structure, would exhibit a fingerprint energy of zero, epitomizing perfect symmetry. Therefore, the process of minimizing fingerprint energy through a tailored fingerprint force-field becomes a strategic approach to optimizing crystal structures towards enhanced symmetry. This methodology not only provides a clear path to achieving higher-symmetry configurations but also offers a nuanced understanding of the underlying symmetry in complex crystal structures.

In order to effectively introduce the fingerprint space during the optimization process, a specific parameterization is incorporated into the mixing strategy, ensuring that the transition towards physically realistic structures is informed by both the exploratory freedom of hyperspace and the the actual features of the target structure.
In the early stages of structural optimization, structures are free to explore in the fingerprint landscape, unconstrained by the typical limitations of normal structural space. This initial phase is crucial for avoiding potential traps and exploring a broader range of conformational possibilities. However, as the optimization process advances, it becomes imperative to guide these structures towards a final configuration that resides entirely in the normal space. To smoothly transition from the expansive exploration of hyperspace to the final, physically realistic structures, the mixing weights are chosen to as follows:
\begin{align}
\begin{split}
&\omega_{\text{fp}}=\mathrm{\Theta}(1-x_{\text{iter}}); \\
&\omega_{\text{real}}=0.5\mathrm{\Theta}(x_{\text{iter}}-1)\left(\sin \left(-\frac{\pi}{2}+9 \pi x_{\text{iter}}^2\right)+1\right),
\end{split}
\end{align}
where $\mathrm{\Theta}$ represents the Heaviside step function, and $x_{\text{iter}}=n_{\text{iter}}/n_{\text{max}}$. Here, $n_{\text{iter}}$ denotes the current step of the structural relaxation, and $n_{\text{max}}$ is a user-defined integer representing the maximum number of structural relaxations before deactivating the fingerprint space.

To implement the structural optimization in the mixed space, it is important to calculate the forces of each atom in guiding the structures to an energetically favorable configuration. The atomic forces on the mixed PES can be obtained from taking the partial derivative of the $3N$ Cartesian coordinates (where $N$ is the total number of atoms in the system) with respect to mixed energy,
\begin{align}
    \boldsymbol{F}_{\text{mixed}}
    &=-\omega_{\text{fp}}\frac{\partial}{\partial \boldsymbol{x}}E_{\text{fp}}-\omega_{\text{real}}\frac{\partial}{\partial \boldsymbol{x}}E_{\text{real}}.
\end{align}

The derivatives of the real energy will be obtained from the DFT or force field calculations, while the derivatives of the fingerprint energy can be defined as follows:
\begin{align}
    \frac{\partial}{\partial x_{k}^{D}}E_{\text{fp}}
    &=2\eta \sum_{i=1}^{N_{\text{at}}} \sum_{j>i}^{N_{\text{at}}}\left(\mathbf{fp}_{i}-\mathbf{fp}_{j}\right)^{\intercal} \cdot\left(\frac{\partial \mathbf{fp}_{i}}{\partial x_{k}^{D}}-\frac{\partial \mathbf{fp}_{j}}{\partial x_{k}^{D}}\right),
    \end{align}
where $x_{k}^{D}$ denotes the $k$-th Cartesian coordinate of the $D$-th atom, and the derivatives of the atomic fingerprint are calculated using the Hellmann-Feynman theorem (see details in \textbf{Supporting Information}). For the crystal structures, the Cauchy stress tensor is calculated using finite difference method, $\sigma_{\alpha \beta}=-\frac{1}{\Omega} \frac{\partial E_{fp}}{\partial \epsilon_{\alpha \beta}}$, where $\epsilon_{\alpha \beta}$ are the elements of the strain tensor. The stress tensor elements $\sigma_{\alpha_\beta}$ are defined as the volume normalized negative strain derivatives of the fingerprint energy.

Our methodology was implemented in \textit{Python} 3, utilizing the Atomic Simulation Environment (ASE) \cite{Hjorth-Larsen-2017} to interface with various calculators to perform energy and force calculations at the level of DFT or force fields.  To demonstrate the effectiveness of our approach, we performed structural optimizations on randomly generated initial structures for both crystal and cluster systems. 

\begin{figure*} 
     \centering
     \begin{subfigure}[b]{0.45\textwidth}
         \centering
         \includegraphics[width=\textwidth]{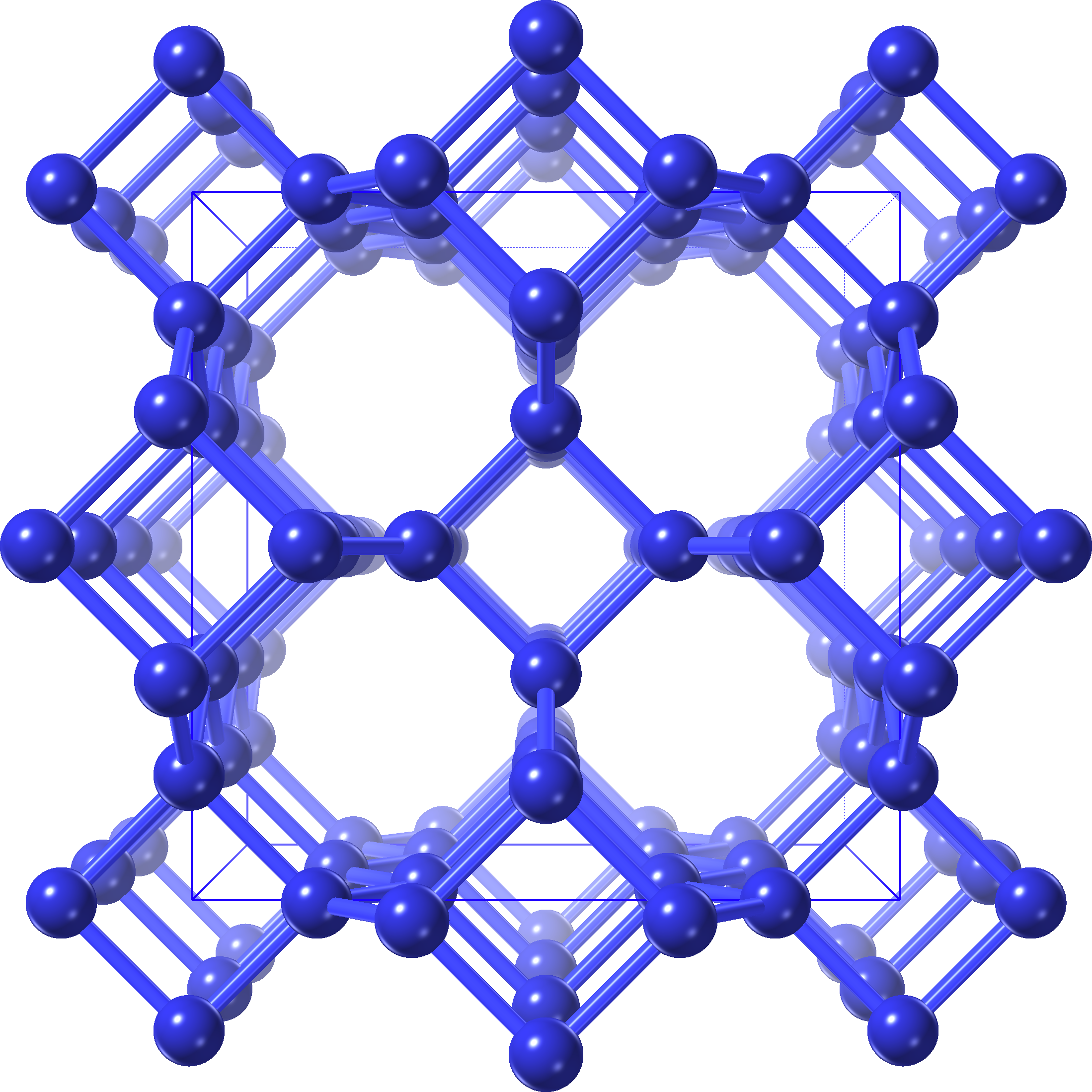}
         \caption{\centering}
         \label{fig:fp_opt_Si_32_DFTB_197}
     \end{subfigure}
     \hfill
     \begin{subfigure}[b]{0.45\textwidth}
         \centering
         \includegraphics[width=\textwidth]{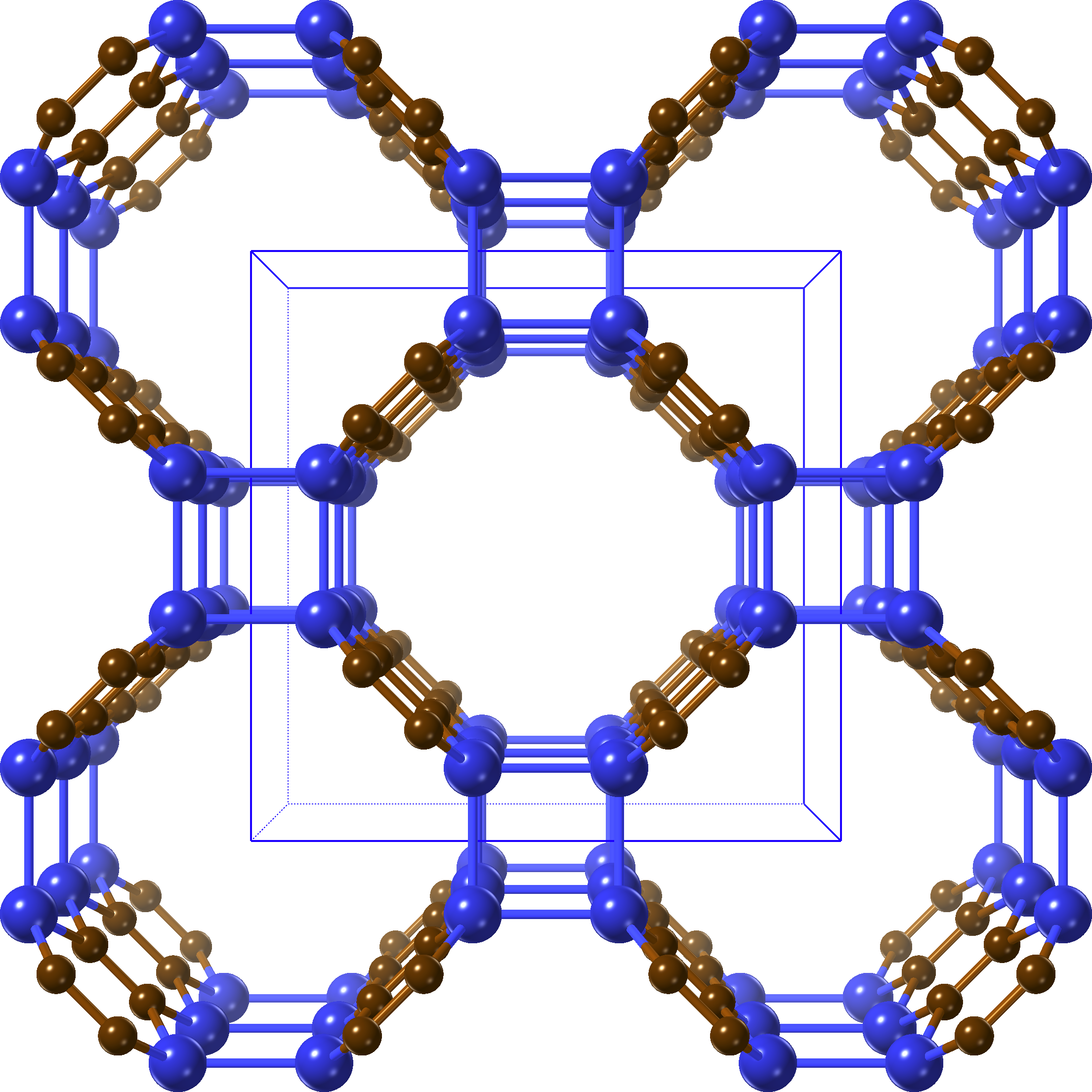}
         \caption{\centering}
         \label{fig:fp_opt_Si_16_C_16_DFTB_33}
     \end{subfigure}
        \caption{(a) Top view (from c-axis) of the relaxed structure ($I4/mmm$-Si$_{32}$) in fingerprint space. 
        The initial structure ($P4nc$-Si$_{32}$) has 8 symmetry operations, but after relaxation in fingerprint space, the structure exhibits 128 symmetry operations. 
      Without fingerprint biasing, the initial structure would be relaxed into a low symmetry  ($P1$). (b) Top view (from c-axis) of the relaxed structure ($P4/mmm$-Si$_{16}$C$_{16}$) in fingerprint space. While the initial structure ($P\Bar{4}2_{1}c$-Si$_{16}$C$_{16}$) has 8 symmetry operations, this fingerprint-relaxed structure has 64 symmetry operations. Without fingerprint biasing, the initial structure will be relaxed into $I\Bar{4}2m$-Si$_{16}$C$_{16}$, which has 16 symmetry operations. Brown and blue spheres represent C and Si atoms, respectively. The blue cubic box denotes the unit cell.}
        \label{fig:fp_opt_structs}
\end{figure*}

For crystals, our focus was on silicon and silicon carbide systems, chosen for their propensity to form complex allotropes and compounds. This leads to intricate and challenging PES landscapes, making them suitable testbeds for our method.   
This complexity arises from the intricate PES and small energy variations relative to their global minima, primarily influenced by $sp^2$ and $sp^3$ hybridization \cite{malonePredictionMetastablePhase2012, oreshonkovNewCandidateReach2020, zwijnenburgExtensiveTheoreticalSurvey2010, mujicaHighpressure2003, ChengInterlayer1988}. We generated initial random structures (300 for each system) using the CALYPSO code \cite{wangCrystalStructurePrediction2010,WANG20122063} without bias towards specific space or point groups. For local optimization, we employed the ASE built-in optimizer, \texttt{FIRE} (\textit{fast inertial relaxation engine}) \cite{PhysRevLett.97.170201} for simultaneous optimization of atomic positions and cell parameters.
The energy and force calculations for the real PES were performed using DFTB$+$ (Density Functional Tight Binding method) \cite{10.1063/1.5143190} with pbc Slater-Koster parameterization set \cite{slaterSimplifiedLCAOMethod1954,sieckShapeTransitionMediumsized2003,raulsStoichiometricNonstoichiometric10101999} and 
0.04 \texttt{k-grid} (equivalent to 0.04$\times2\pi$\AA$^{-1}$ $k$-point meshes). Figure \ref{fig:fp_opt_structs} shows the comparison of the distribution of local minima obtained through  optimization on the real PES with those identified using the mixed PES approach. We can find that the fingerprint space biasing technique effectively shifts the energy landscape, promoting the emergence of high-symmetry structures. For both the Si$_{32}$ and Si$_{16}$C$_{16}$ systems, there is a significant density of high-symmetry structures at lower enthalpy levels when the optimization is guided by the fingerprint space, indicating the probability of finding energetically more stable  structures is significantly enhanced. 
In addition to identifying lower-energy structures, our method also successfully identified high-symmetry structures that were not found in the optimization using the real PES. For example, the $I4/mmm$-Si$_{32}$ structure (FIG. \ref{fig:fp_opt_Si_32_DFTB_197}) and the $P4/mmm$-Si$_{16}$C$_{16}$ structure (FIG. \ref{fig:fp_opt_Si_16_C_16_DFTB_33}) were discovered during the local optimization when relaxed on the mixed PES. This indicates that our method can successfully identify high-symmetry structures that might otherwise be missed or underrepresented in conventional optimizations.

\begin{figure*} 
     \centering
     \begin{subfigure}[b]{0.45\textwidth}
         \centering
         \includegraphics[width=\textwidth]{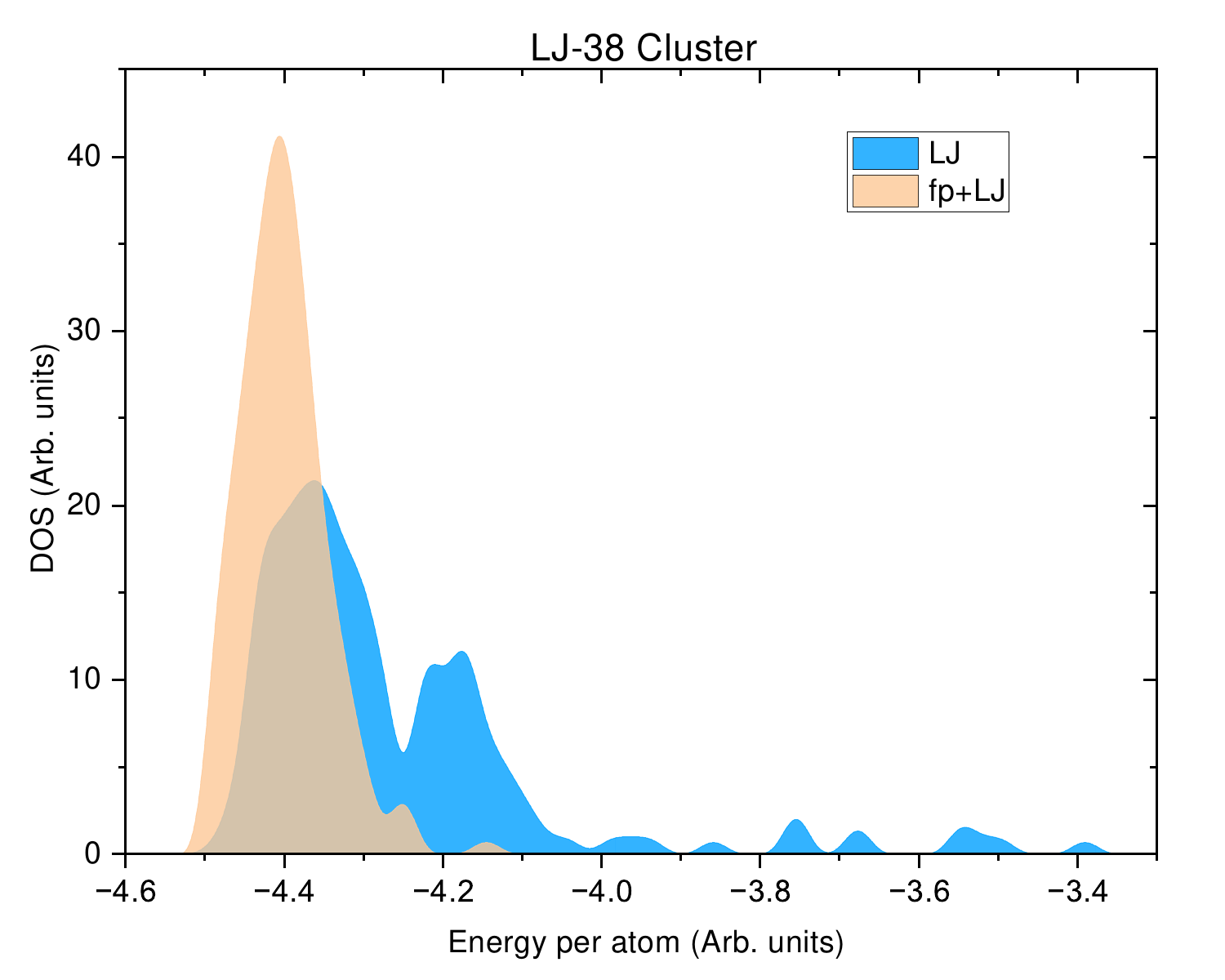}
         \caption*{}
         \label{fig:LJ38_clusters}
     \end{subfigure}
     \hfill
     \begin{subfigure}[b]{0.45\textwidth}
         \centering
         \includegraphics[width=\textwidth]{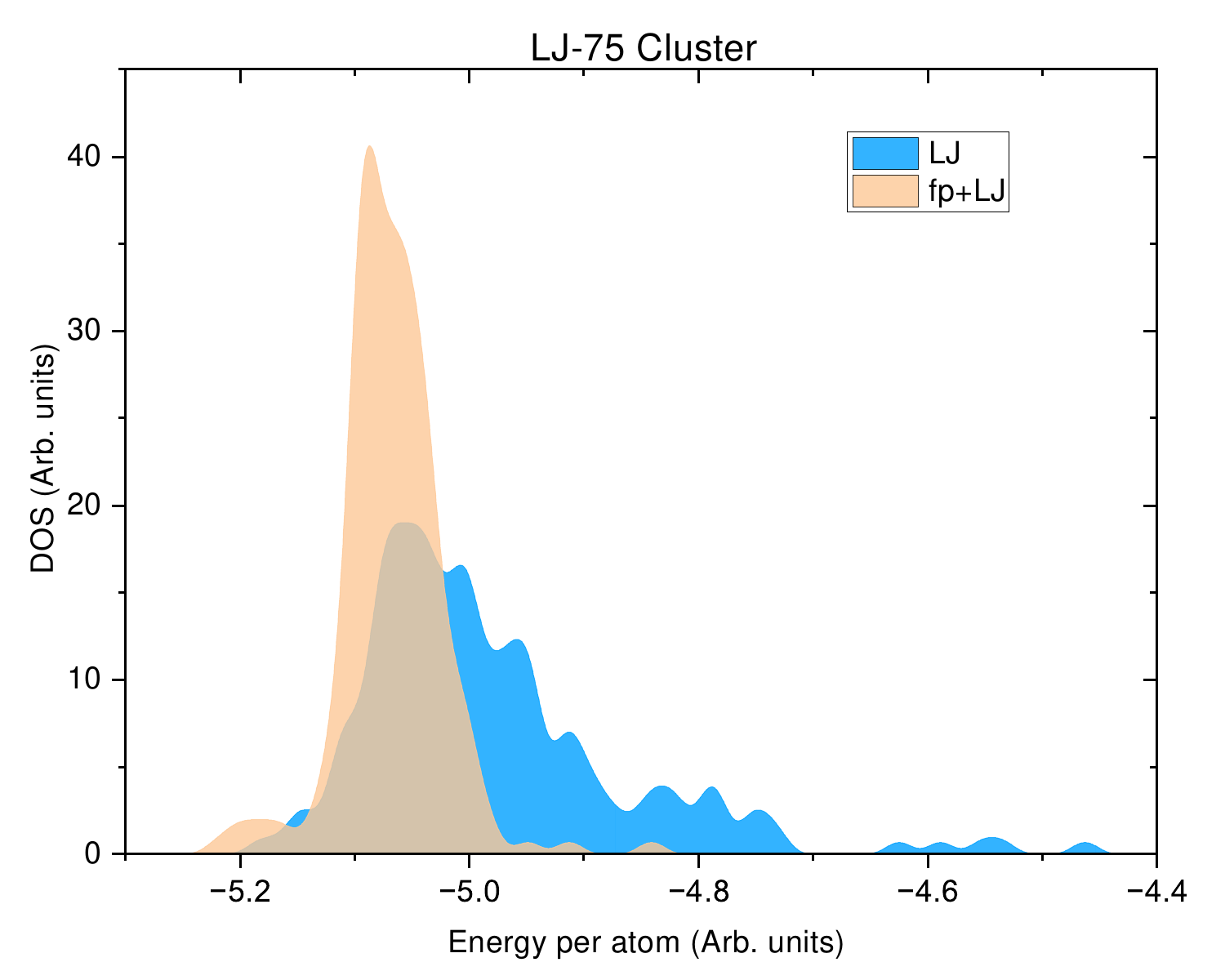}
         \caption*{}
         \label{fig:LJ75_clusters}
     \end{subfigure}
     \\
     \begin{subfigure}[b]{0.45\textwidth}
         \centering
         \includegraphics[width=\textwidth]{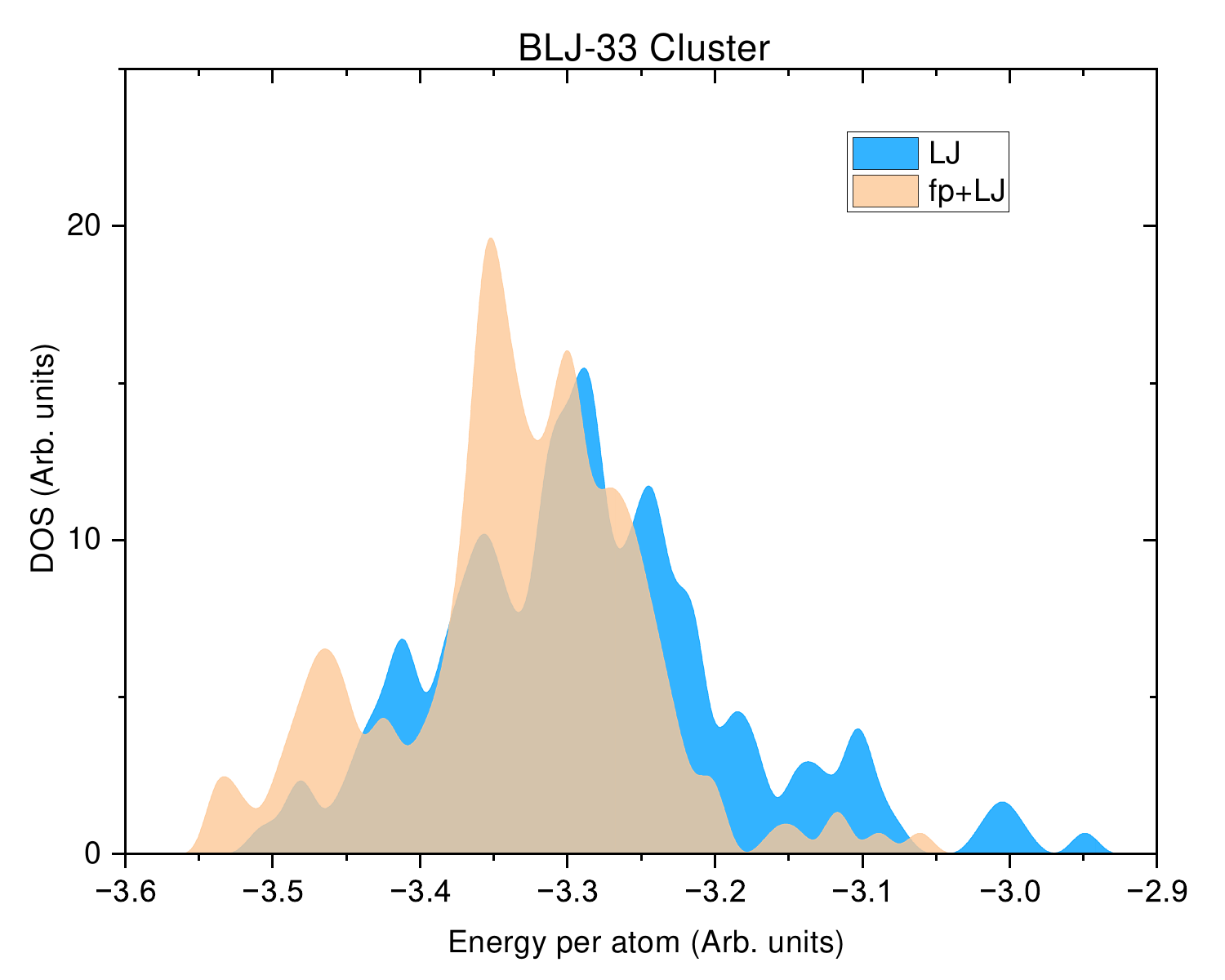}
         \caption*{}
         \label{fig:BLJM33_cluster}
     \end{subfigure}
     \hfill
     \begin{subfigure}[b]{0.45\textwidth}
         \centering
         \includegraphics[width=\textwidth]{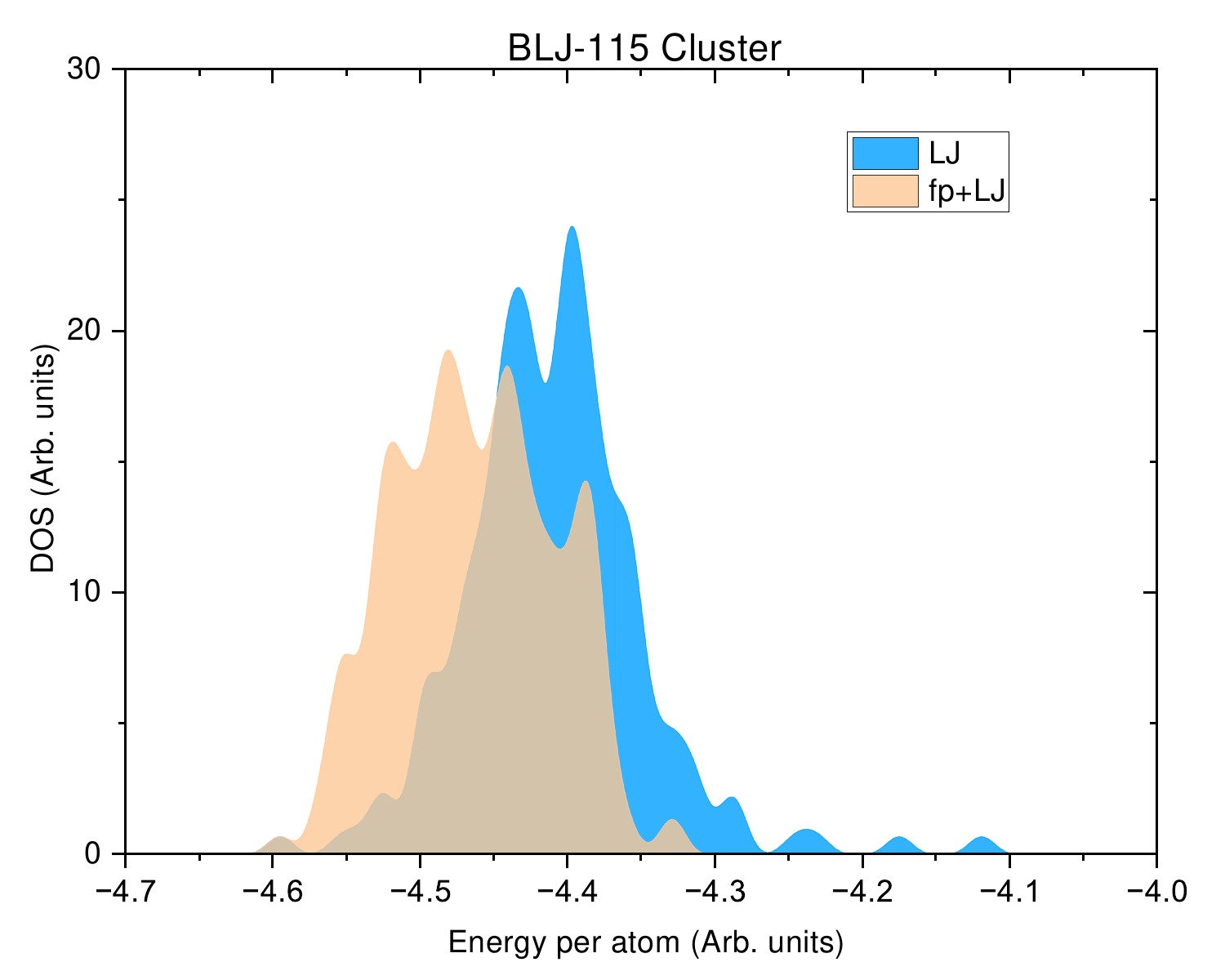}
         \caption*{}
         \label{fig:BLJM115_clusters}
     \end{subfigure}
        \caption{Density of states (DOS) of the energy distribution of the optimized random-generated clusters with (in salmon) or without (in blue) introducing fingerprint space. For binary LJ clusters, we choose the chemical formula as A$_{13}$B$_{20}$ in the BLJ-33 case and and A$_{55}$B$_{60}$ in the BLJ-115 case.}
        \label{fig:fp_cluster_hist}
\end{figure*}

We also performed structural optimizations on Lennard-Jones (LJ) and binary Lennard-Jones (BLJ) clusters, which are commonly used as benchmark systems in structure optimization methods due to their well-defined PES landscapes. Our method was first tested on LJ$_{38}$ and LJ$_{75}$ clusters, known for their nontrivial double-funnel energy landscapes \cite{doyeDoublefunnelEnergyLandscape1999}. 
For the BLJ clusters, the parameters $\sigma_{AA} = 1.0$, $\sigma_{BB} = 0.8$, $\epsilon_{AA} = 1.0$, $\epsilon_{BB} = 0.64$ are chosen, with $\sigma_{AB}=0.5\times\left(\sigma_{AA}+\sigma_{BB}\right)$ and $\epsilon_{AB}=\sqrt{\epsilon_{AA}\epsilon_{BB}}$.\cite{mravlakStructureDiagramBinary2016,yeGlobalOptimizationBinary2011}
A total of 200 random configurations were generated for each cluster using the CALYPSO code,  and the energy and force calculations were performed using the ASE built-in calculator \cite{Hjorth-Larsen-2017}. Figure \ref{fig:fp_cluster_hist} shows the comparison of  the density of states (DOS) as a function of the energy per atom for different cluster optimizations. We can find that our method effectively shifts the energy landscape in a way that promotes the emergence of low-energy configurations. For both the LJ and BLJ systems, there is a significant density of low-energy structures when the search is guided by the fingerprint space, indicating an enhanced probability of finding energetically more stable structures.

\begin{figure*} 
     \centering
     \begin{subfigure}[b]{0.45\textwidth}
         \centering
         \includegraphics[width=\textwidth]{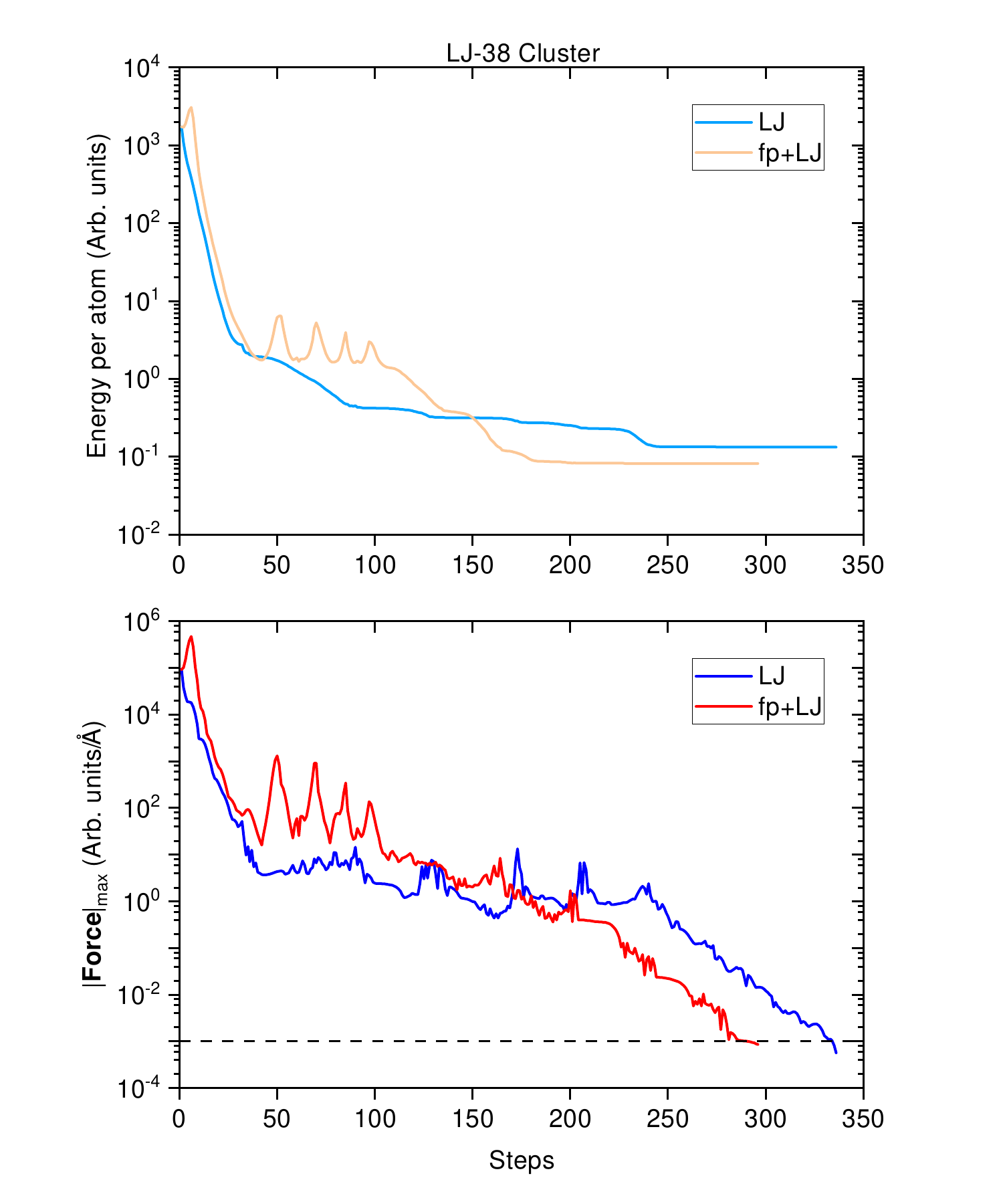}
         \caption*{}
         \label{fig:LJ_38_traj}
     \end{subfigure}
     \begin{subfigure}[b]{0.45\textwidth}
         \centering
         \includegraphics[width=\textwidth]{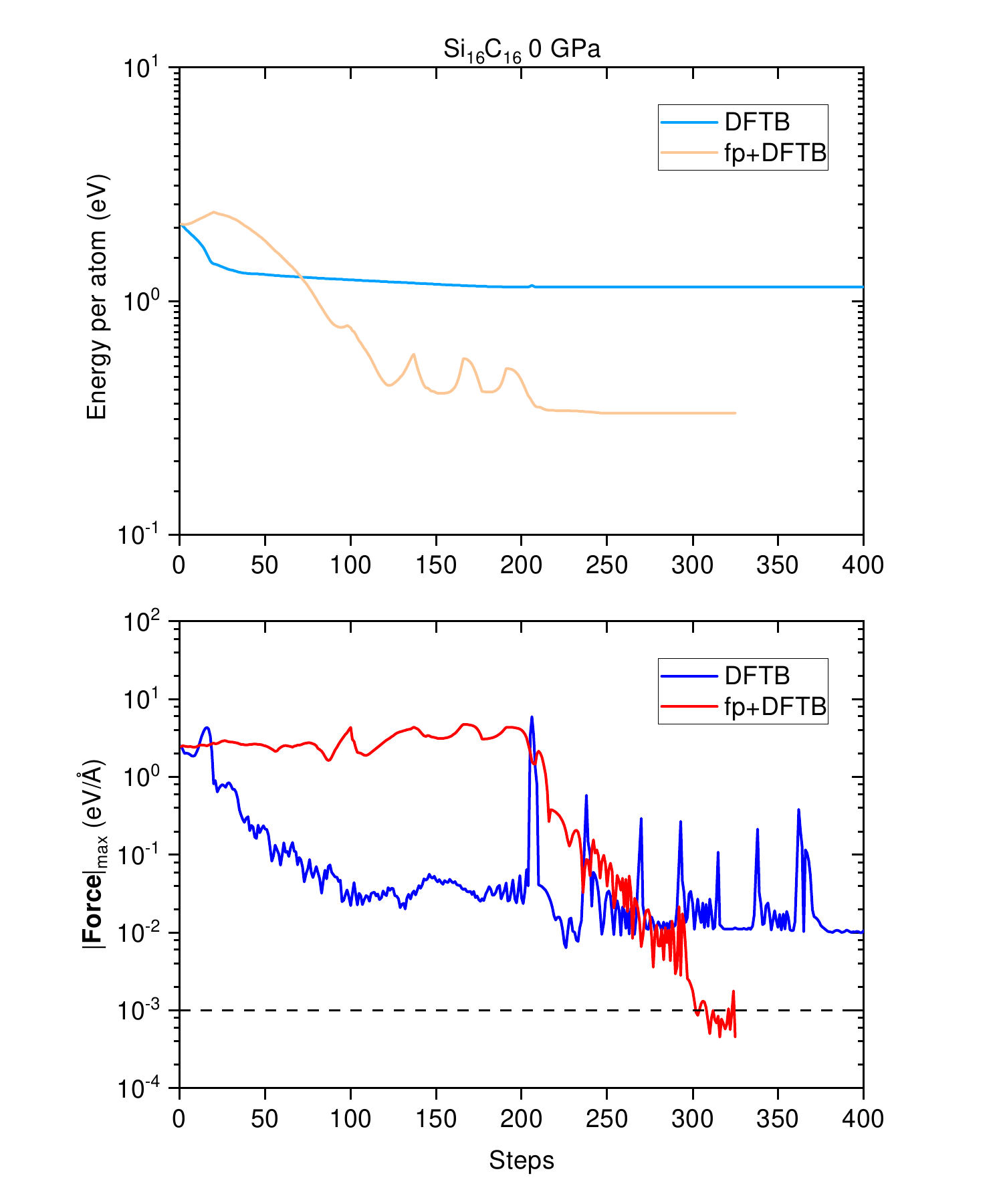}
         \caption*{}
         \label{fig:Si_16_C_16_DFTB_traj}
     \end{subfigure}
        \caption{Trajectory plots for LJ-38 cluster (left) and Si$_{16}$C$_{16}$ crystal (right) during local optimization process. The top panel shows the energy at each local relaxation step. The bottom panel displays the maximum component of the forces of all the atoms in the system at each local relaxation step. The force convergence criteria (dashed line in the bottom panel) is set at $1\times10^{-3}$ in the corresponding units.}
        \label{fig:traj_plot}
\end{figure*}

To further investigate the efficiency of our method, we plot the trajectories (FIG. \ref{fig:traj_plot}) for LJ-38 and Si$_{16}$C$_{16}$ in the local optimization.  Utilizing the FIRE (\textit{fast inertial relaxation engine}) \cite{PhysRevLett.97.170201} optimizer, we observe a notable acceleration in the energy and force convergence when the relaxation process is conducted within the mixed space. This is particularly evident in the case of the silicon carbide system, where the convergence is markedly swift. For the Si$_{16}$C$_{16}$ system, we encounter a funnel-like energy landscape which typically induces sluggish convergence. In our case, the optimization in the normal PES failed to converge within 5000 steps. 
However, the integration of our fingerprint space dramatically alters this dynamic. It becomes evident that within the initial 100 steps, the optimizer adeptly surpasses substantial energy barriers. This capability is crucial, as it allows the system to access lower energy minima in subsequent geometry relaxation steps. Our approach thus serves as a testament to the power of incorporating hyper space strategies, which promise to advance the field of computational material science.

In conclusion, we have developed a new method to utilize the intrinsic symmetry of atomic structures to enhance the efficiency of local optimization. By integrating a fingerprint energy derived from the symmetry of local atomic environments into the potential energy surface, we have developed an approach that guides the optimization process towards high-symmetry, low-energy structures. This implicit symmetry bias, embedded in the hyperdimensional optimization space, has been shown to  accelerate the discovery of energetically favorable configurations in both crystal and cluster systems, thereby streamlining the search for new materials.
The benchmark results obtained from the application of our method to silicon, silicon carbide, and Lennard-Jones cluster systems underscore its effectiveness. Notably, the emergence of high-symmetry structures at lower energy levels and the rapid convergence of energy and forces during optimization highlight the potential to reveal previously underexplored regions of the potential energy landscape. 
Overall, our method represents a step forward in the structure prediction field, offering an efficient tool for the exploration of complex potential energy surfaces. As we continue to refine and apply this approach to a wider range of materials, we anticipate that it will become an important component of the toolkit in computational material science.

\begin{figure*}
     \centering
     \begin{subfigure}[b]{0.45\textwidth}
         \centering
         \includegraphics[width=\textwidth]{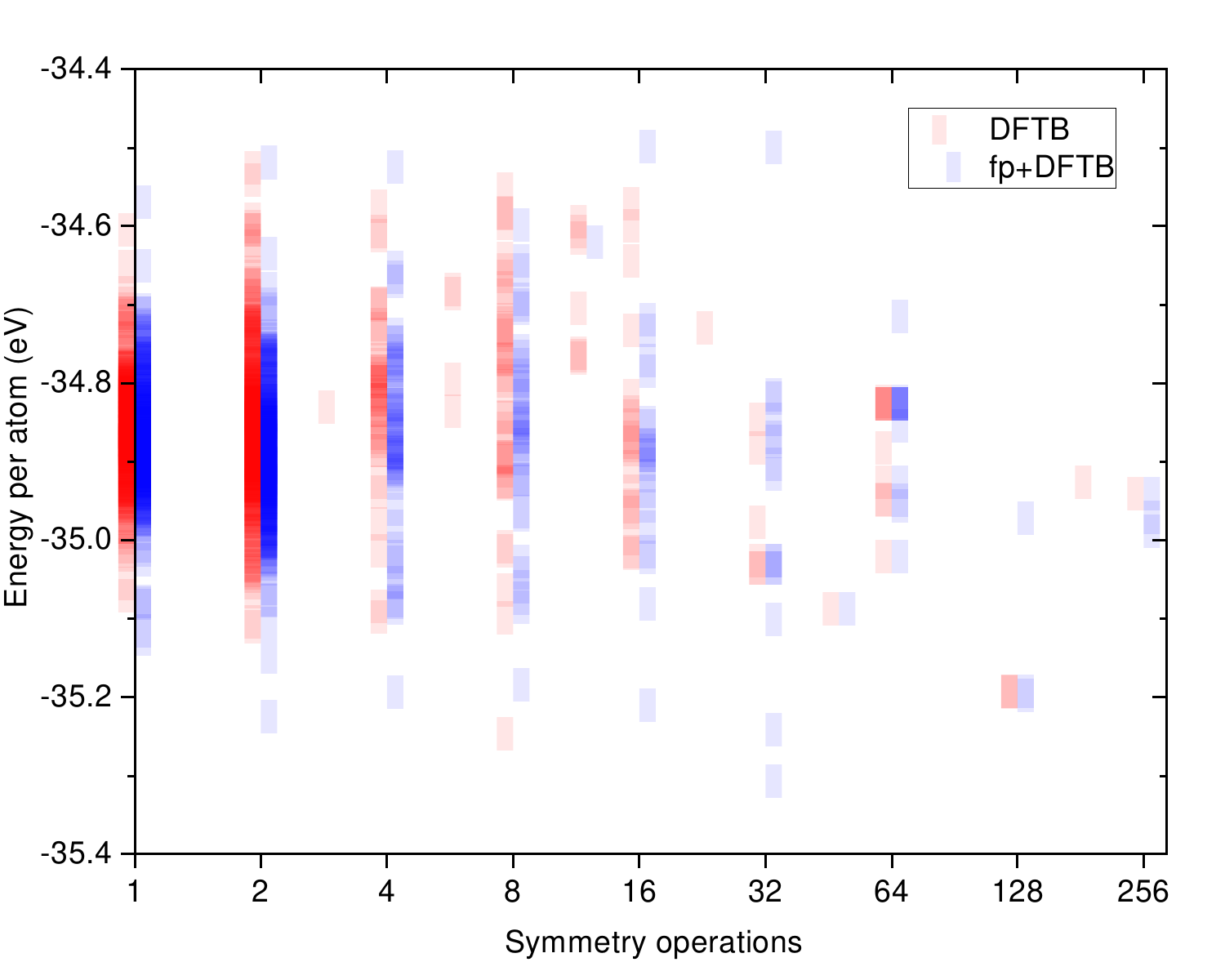}
         \caption{\centering}
         \label{fig:Si_32_H_VS_sym_DFTB_0GPa}
     \end{subfigure}
     \hfill
     \begin{subfigure}[b]{0.45\textwidth}
         \centering
         \includegraphics[width=\textwidth]{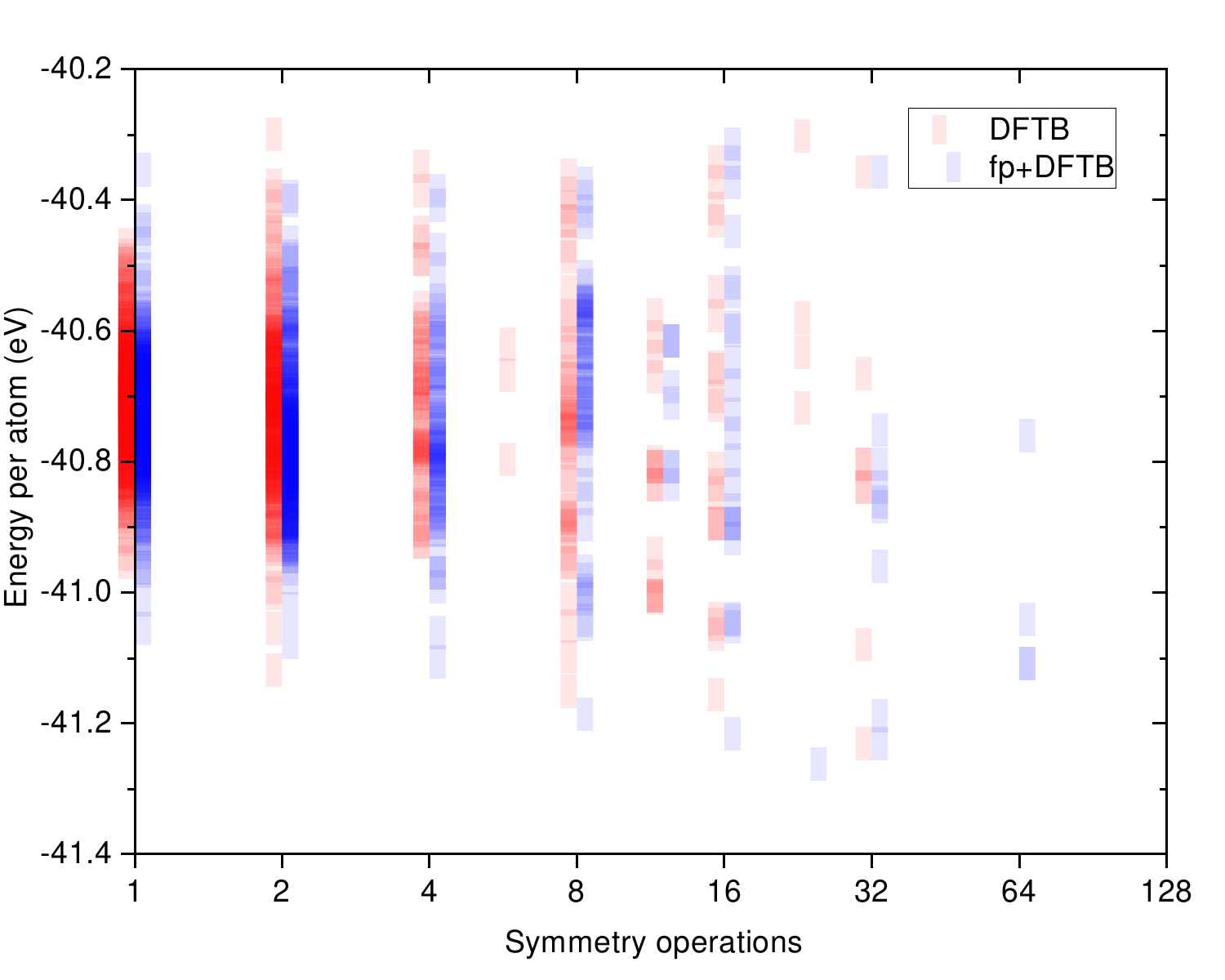}
         \caption{\centering}
         \label{fig:Si_16_C_16_H_VS_sym_DFTB_0GPa}
     \end{subfigure}
        \caption{Comparison of the distribution of local minima optimized on real PES (red) and on fingerprint mixed PES (blue) for (a) Si$_{32}$ crystal and (b) Si$_{16}$C$_{16}$ crystal at ambient conditions. Symmetry, plotted along the x-axis, is measured by the number of symmetry operations that leave the structure invariant. The opacity of the colored box indicates the relative abundance of meta-stable structures corresponding to a given symmetry and enthalpy level.}
        \label{fig:H_VS_sym}
\end{figure*}

\begin{acknowledgement}

This work was supported by the National Science Foundation, Division of Materials Research (NSF-DMR) under Grant No. 2226700, and startup funds of the office of the Dean of SASN of Rutgers University-Newark. The authors acknowledge the Office of Advanced Research Computing (OARC) at Rutgers, The State University of New Jersey,
for providing access to the Amarel cluster and associated research computing resources that have contributed to the results reported here. 

\end{acknowledgement}





\bibliography{mainbib}

\end{document}